\documentclass[prb,twocolumn,showpacs,superscriptaddress]{revtex4}

\usepackage{graphicx}
\usepackage{dcolumn}
\usepackage{amsmath}

\begin{document}

\title[MD simulations of laser ablation of silicon]{\textbf
Molecular-dynamics thermal annealing model\\ of laser ablation of silicon}

\author{\firstname{Patrick} \surname{Lorazo}}
   \email[E-mail: ]{lorazop@magellan.umontreal.ca}
   \affiliation{D\'{e}partement de G\'{e}nie Physique et de G\'{e}nie des
      Mat\'{e}riaux et Groupe de Recherche en Physique et Technologie des
      Couches Minces (GCM), \'{E}cole Polytechnique de Montr\'{e}al, Case
      Postale 6079, Succursale Centre-Ville, Montr\'{e}al, Qu\'{e}bec, Canada,
      H3C 3A7}
   \affiliation{D\'{e}partement de Physique et Groupe de Recherche en Physique
      et Technologie des Couches Minces (GCM), Universit\'{e} de Montr\'{e}al,
      Case Postale 6128, Succursale~Centre-Ville, Montr\'{e}al, Qu\'{e}bec,
      Canada H3C 3J7}
\author{\firstname{Laurent J.} \surname{Lewis}}
   \email[To whom correspondence should be addressed; e-mail:]
      {laurent.lewis@umontreal.ca}
   \affiliation{D\'{e}partement de Physique et Groupe de Recherche en
      Physique et Technologie des Couches Minces (GCM), Universit\'{e} de
      Montr\'{e}al, Case Postale 6128, Succursale~Centre-Ville, Montr\'{e}al,
      Qu\'{e}bec, Canada H3C 3J7}
\author{\firstname{Michel} \surname{Meunier}}
   \email[E-mail: ]{meunier@phys.polymtl.ca}
   \affiliation{D\'{e}partement de G\'{e}nie Physique et de G\'{e}nie des
      Mat\'{e}riaux et Groupe de Recherche en Physique et Technologie des
      Couches Minces (GCM), \'{E}cole Polytechnique de Montr\'{e}al, Case
      Postale 6079, Succursale Centre-Ville, Montr\'{e}al, Qu\'{e}bec,
      Canada, H3C 3A7}


\date{\today}

\begin{abstract}
A molecular-dynamics thermal annealing model is proposed to investigate the
mechanisms involved in picosecond pulsed laser ablation of crystalline
silicon. In accordance with the thermal annealing model, a detailed
description of the microscopic processes which result from the interaction of
a 308 nm, 10 ps, Gaussian pulse with a Si(100) substrate has been embedded
into a molecular-dynamics scheme. This was accomplished by explicitly
accounting for carrier-phonon scattering and carrier diffusion. Below the
predicted threshold fluence for ablation, $F_{th}=0.25\ \text{J/cm}^{2}$, a
surface breathing mode indicates that the solid restores internal equilibrium
by the generation of pressure waves. Above $F_{th}$, our simulations reveal
that matter removal is triggered by subsurface superheating effects: intense
heating of the material leads to the thermal confinement of the
laser-deposited energy. As a result, the material is overheated up to a temperature
corresponding to the critical point of silicon and a strong pressure gradient builds up
within the absorbing volume. At the same time, diffusion of the carriers into the bulk
leads to the development of a steep temperature gradient beneath the surface. Matter
removal is subsequently driven by the relaxation of the pressure gradient:
large pieces --- several atomic layers thick --- of molten material are
expelled from the surface with initial axial velocities of $\sim 1000\ \text{m/s}$,
their ejection following the nucleation of voids beneath the surface.
\end{abstract}


\maketitle


\section{Introduction\protect\\}
\label{sec:intro}

The use of photons for the controlled removal of matter, i.e., laser
ablation,\cite{chase94,haglund98} has given rise to a large number
of applications for materials processing and growth; in microelectronics,
examples include surface micromachining, surface cleaning, and the pulsed
laser deposition (PLD)\cite{chrisey} of thin films, which allows a wide
variety of materials to be grown.\cite{chrisey,belouet96} In spite of this, a
complete picture of the mechanisms underlying ablation for a broad range of
pulse durations and wavelengths, and for such diverse materials as metals,
polymers and semiconductors, is still lacking.

The great complexity of phenomena involved in laser ablation, determined by
the coupling of the laser pulse with the optical, elastic and thermal
properties of the material, has been an obstacle in understanding. On a
mesoscopic scale, the processes brought about by ultrashort --- femtosecond
--- pulses have seldom been addressed.\cite{linde98,linde00,schmidt00} On
the other hand, evidence of a competition between
photomechanical\cite{jacques95,zhigilei00} and
photothermal\cite{miotello95,zhigilei00} mechanisms has begun to emerge in
the light-induced removal of matter with picosecond and nanosecond pulses. In
the \textit{stress confinement} regime,\cite{jacques95,zhigilei00} ejection
of matter proceeds from photomechanical effects: energy is deposited in the
material in a time shorter than that needed for the generation of acoustic
waves, and ablation is induced by the relaxation of the pressure gradient
that builds up within the absorbing volume. In the \textit{thermal
confinement} regime,\cite{zhigilei00} energy is supplied to the irradiated
volume at a rate faster than that at which it can be pumped out by thermal
diffusion. Consequently, the material is rapidly brought near its critical
point as matter removal is driven by photothermal effects:\cite{miotello95}
the metastable superheated liquid undergoes a prompt transition to a gaseous
phase consisting, to a large extent, of molten micron-sized droplets. This
mechanism is known as explosive boiling or phase explosion;\cite{miotello95}
other photothermal processes include vaporization and normal boiling (see
Ref.~\onlinecite{kelly96} and references therein). Finally, there have been
repeated reports\cite{dabby72,bhattacharya91,singh90,craciun98,craciun97} of
a fourth process known as subsurface superheating (SSSH). In this scenario,
heating of the material causes evaporation as a result of which a steep
temperature gradient develops beneath the surface. Matter removal then takes
place in a manner similar to a phase explosion following the relaxation of
the pressure build-up in the bulk. It is not clear whether this mechanism is
real as conflicting views can be found in the
literature.\cite{miotello95,kelly96,craciun97,craciun98}

While it is generally assumed that the laser sputtering of metal surfaces
involves phase explosion mechanisms (see, e.g., Ref.\
\onlinecite{bennett95}), both photomechanical and photothermal effects have
been shown to be operative in organic solids.\cite{zhigilei00} In
semiconductor materials such as silicon, however, understanding of matter
removal processes induced by picosecond pulses has yet to come. In the
present study, we give evidence of a SSSH effect in the laser ablation of
silicon. This is achieved by carrying out detailed molecular-dynamics (MD)
simulations of the interaction of a picosecond laser pulse with a Si(100)
substrate, explicitly accounting for carrier-phonon scattering and carrier
diffusion. It is assumed that the thermal annealing model (TAM)\cite{wood84}
is valid so that thermal processes only are operative. In our computer code,
`MADTAM', we use the Stillinger-Weber (SW)\cite{stillinger85} potential to
model the atomic interactions; preliminary results from an earlier version
of our model can be found in Refs. \onlinecite{lorazo00} and \onlinecite{lorazo00_2}.

Anticipating our results, we find that, below the (predicted) threshold
fluence for ablation, $F_{th}$, the solid restores internal equilibrium by
the generation of pressure waves. Above $F_{th}$, our simulations reveal that
matter removal is triggered by SSSH effects: intense heating of the material
leads to the thermal confinement of the laser-deposited energy and a steep
temperature gradient beneath the surface. Interestingly, the latter is not
caused by the evaporation of atoms from the outer surface but, rather,
originates from the process of carrier diffusion into the bulk. As a result,
a strong pressure gradient builds up $\sim 30\ \text{nm}$ below the surface as
the material is overheated up to a temperature corresponding to the critical point of
silicon. Ablation is subsequently initiated by the relaxation of the pressure gradient:
large pieces --- several atomic layer thick --- of molten material are expelled from the
surface, a consequence of the nucleation of voids in the bulk. Before
discussing these results, however, we give a detailed description of our
model.

\section{The model\protect\\}
\label{sec:model}

\begin{figure}
\includegraphics*[width=7.5cm,height=9.05cm]{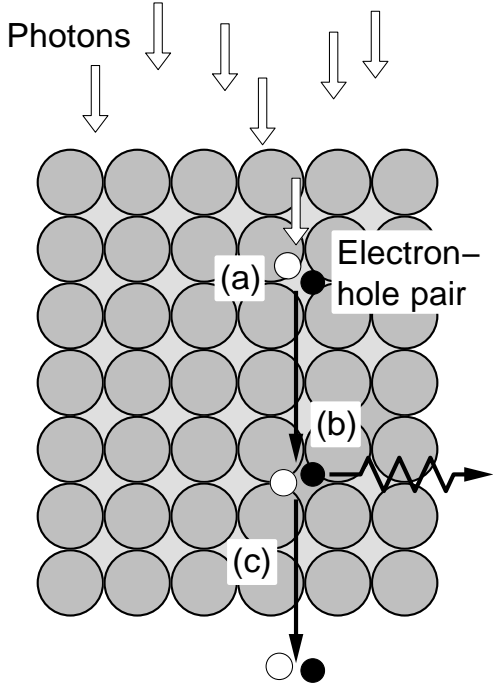}
\caption{Generation of an $e$-$h$ pair and the main subsequent relaxation
mechanisms in MADTAM.\ (a) creation of an $e$-$h$ pair upon absorption of a
photon; (b) scattering of the electron or hole by a phonon; (c)
unidirectional diffusion of the $e$-$h$ pair into the bulk along the $z$
axis.}
\label{cartoon}
\end{figure}

The processing of silicon surfaces with femtosecond laser pulses involves
ultrafast, non-thermal, phenomena which cannot be accounted for by the SW
potential. For this reason, we remain within the picosecond regime where
thermal mechanisms are involved; this is discussed in Sec.~\ref{preliminary}.
A description of the computational method, i.e., molecular dynamics, follows
in Sec.~\ref{md}. Finally, the generation of electron-hole (\textit{e-h})
pairs by photons, as well as their relaxation through carrier-phonon
scattering and carrier diffusion, depicted in Fig.~\ref{cartoon}, are
detailed in Sec.~\ref{subsec:target} to~\ref{subsec:relax}.

\subsection{Preliminary considerations\protect\\}
\label{preliminary}

Various computational techniques allow the interaction of light with matter
to be simulated.\cite{lietoila82,vandriel87,
pronko95,seifert98,garrison85,vertes95,campbell98,zhigilei97,zhigilei98,zhigilei00,lorazo00}
Among them, MD appears to be a most suitable approach for addressing the
numerous, complex processes involved in the laser ablation and desorption of
matter: no hypotheses are needed to account for the various phenomena and
atomic motion can be followed, in real time, provided that the interactions
between the atoms are correctly modeled.\cite{allen87,smith97} However,
because laser ablation involves large length and time scales, and thus a
great computational effort, attempts involving classical MD are
scarce.\cite{garrison85,vertes95,campbell98,zhigilei97,zhigilei97_2,zhigilei98,zhigilei00,lorazo00}
In order to extend the range of MD simulations, Zhigilei \textit{et al.}
proposed a breathing-sphere model appropriate to the study of matter removal
in organic solids.\cite{zhigilei97} In this work, the phase explosion due to
overheating and the laser-induced pressure build-up were identified as the
key processes responsible for ablation in the regimes of thermal and stress
confinements, respectively.\cite{zhigilei97,zhigilei00}

Another MD study of laser ablation was performed by Herrmann \textit{et
al.},\cite{campbell98} who examined the interaction of femtosecond laser
pulses with silicon; a truncated version of the SW potential was used to
account for the rupture of interatomic bonds following the absorption of
photons. The predicted values of the threshold fluence for ablation thus
obtained were at least an order of magnitude higher than those observed
experimentally.\cite{cavalleri99,cavalleri98} This discrepancy is likely due
to the inability of the SW potential to describe the non-thermal mechanisms
involved in the interaction of femtosecond pulses with silicon. Indeed, it is
generally agreed\cite{linde97,silvestrelli96,linde00_2} that there exists, for silicon,
a critical carrier density $n_{c}\sim 10^{22}\ \text{cm}^{-3}$ separating two
distinct regimes:

(i) With picosecond and nanosecond pulses, carrier densities remain below the
critical value and the relaxation is thermal, i.e., the carriers relax by
transferring their kinetic energy to the lattice by the spontaneous emission
of optical phonons in a characteristic time of $\sim 1\ \text{ps}$. In this
context, structural modifications such as melting and ablation are observed
on a $\sim 100-1000\ \text{ps}$ time scale. This model, known as TAM, is suitable
for laser pulses down to approximately 10~ps.\cite{wood84}

(ii) With femtosecond pulses, on the other hand, the number of free carriers
may exceed $n_{c}$ and ultrafast melting proceeds within $\sim 1\ \text{ps}$
by the collapse of the lattice.\cite{stampfli94,silvestrelli97} Because
melting occurs before the carriers have time to relax through the emission of
optical phonons, the relaxation is non-thermal. This model, first proposed by
Van~Vechten \textit{et al.},\cite{vanvechten79} is known as the plasma
annealing model (PAM).\cite{murakami84}

Furthermore, if the light intensity at the surface exceeds $\sim10^{4}\
\text{GW/cm}^{2}$, the generation of a high number of energetic carriers
through a two-photon absorption mechanism leads to the complete ionization of
the material by avalanche ionization and/or field ionization: a very dense
($n\geq10^{23}\ \text{cm}^{-3}$) and hot ($\sim10^{6}\ \text{K}$) $e$-$h$ gas is
generated near the surface.\cite{schuler96,private} One then speaks of a
solid-to-plasma transition as the material is carried away by the expanding
plasma.

Here, because we make use of the SW potential, it is assumed that thermal
processes only are operative; this restricts us to 10~ps or longer pulses.
Care is taken to avoid carrier densities in excess of $n_{c}$; in this
context, the intensity remains well below $\sim10^{4}\ \text{GW/cm}^{2}$.

\subsection{Molecular dynamics\protect\\}
\label{md}

The MD technique has been discussed extensively in the literature (see, for
instance, Ref.~\onlinecite{smith97}). Briefly, the set of $3N$ Newton
equations of motion, where $N$ is the number of atoms in the system, are
solved from the knowledge of the interatomic forces. If
$\textbf{\textit{r}}_{i}$, $\textbf{\textit{v}}_{i}$, and
$\textbf{\textit{a}}_{i}$ are the position, velocity and acceleration of the
$i$th atom, respectively, the time evolution of the system is obtained by
integrating, at discrete time steps, the following differential equations:
\begin{eqnarray}
   \frac{\partial \textbf{r}_{i}}{\partial t}&=&\textbf{v}_{i}\ , \nonumber \\ \nonumber \\
   \frac{\partial \textbf{v}_{i}}{\partial t}&=&\textbf{a}_{i}=-\frac{1}{m_{i}}\left[\frac{\partial
   \textit{U}(\textbf{r}_{1},...,\textbf{r}_{N})}
   {\partial \textbf{r}}\right]_{\textbf{r}=\textbf{r}_{i}}\ ,
\end{eqnarray}
given a set of initial positions and velocities
\{$\textbf{r}^{\text{0}}_{i}$,$\textbf{v}^{\text{0}}_{i}$\}. In the above equations,
$m_{i}$ is the mass of the $i$th atom and $U$ the interatomic potential.
Thermodynamic quantities such as temperature and pressure can then be
computed in a straightforward manner.\cite{allen87}

All calculations were performed using the program {\texttt groF}, a
general-purpose MD code for bulk and surfaces developed by one of the authors
(LJL). The original form of the SW potential was used to compute the forces
between the silicon atoms.\cite{stillinger85} The equations of motion were
solved using the velocity Verlet algorithm.\cite{allen87} The MD time step
was set to $\Delta t=0.5\ \text{fs}$ for all simulations.

\subsection{Target and laser pulse\protect\\}
\label{subsec:target}

Simulations are carried out for a supercell with approximate dimensions
$3\times3\times60\ \text{nm}^{3}$ (containing a total of 31680 atoms)
representing a small volume of a Si(100) target located at the center of the
laser pulse.

In order to recreate the thermal and structural constraints of a macroscopic
crystal, the supercell is first repeated in the lateral ($x$ and $y$)
directions using periodic boundary conditions (PBC). The thermal constraint
is imposed by coupling a few monolayers of atoms at the bottom of the
supercell to a heat reservoir; this is done by renormalizing the velocities
to an appropriate Maxwell-Boltzmann (MB) distribution. These atoms also serve
to minimize the reflection of pressure waves at the bottom of the supercell.
This latter phenomenon affects the results by, in some cases, adding an
artificial contribution to ablation. Though other techniques were proposed to
attenuate such pressure waves,\cite{zhigilei99} the method described above
gives satisfactory results. Additionally, a few monolayers of atoms attached
to their equilibrium positions are placed beneath the heat reservoir in order
to mimic a semi-infinite crystal. The crystal is assumed to be initially
perfect.

The simulation of a 10~ps laser pulse, Gaussian in time as well as in
the $x$ and $y$ directions and of macroscopic width, is accomplished assuming
that any spatial variation of the irradiance can be neglected. This is valid
because the laser spot diameter is assumed to be much larger than the region
under study. The irradiance is thus, in effect, spatially constant over the
infinite $x$-$y$ plane by virtue of the PBC.

The relatively large number of photons contained in a pulse, typically a few
tens of thousands, ensures a uniform spatial distribution of the energy at
the surface. The temporal Gaussian distribution is simulated by the
successive arrival of planes of photons spanning the entire surface of the
supercell, the number of which being determined by the instantaneous
irradiance; they are separated in time by intervals ranging from $\Delta t$
to typically $10\times\Delta t$.

\subsection{Absorption of light\protect\\}
\label{subsec:absorp}

The absorption of light in silicon proceeds by the excitation of valence
electrons via intraband and interband transitions, the frequency of which
depends on several parameters, such as the density of free carriers and the
wavelength and intensity of the laser pulse. The absorption of photons of
energy $h\nu$ larger than the bandgap $E_{G}$ can induce
three types of optical transitions: (i) An interband transition following the
absorption of a single photon by a valence-band electron; a bond is broken
and the electron is promoted to the conduction band, leaving a hole behind.
The newly created pair shares an energy $\sim(h\nu-E_{G})$,
each carrier receiving an initial kinetic energy determined by the set of
selection rules. (ii) A nonlinear, two- (or more) photon, interband
transition, occurring when $n \geq 2$ photons are simultaneously absorbed.
If $nh\nu\geq\phi$, where $\phi=4.85\ \text{eV}$ is the silicon work
function,\cite{handbook} the electron can be photoemitted; the condition
$nh\nu<\phi$, i.e., $\lambda\gtrsim255\ \text{nm}$ when $n=1$, thus ensures
that no photoemission takes place. (iii) An intraband transition, or
free-carrier absorption, according to which the photon is absorbed by an
electron already in the conduction band.

Following their creation, the carriers, because of momentum and energy
conservation rules, occupy thin energy shells and a small volume of
\textbf{k} space. In a characteristic time $\tau_{eh}$ of typically a few
tens of fs, the hole and electron subsystems achieve, through electron-hole
scattering, a quasi-equilibrium state described by a Fermi-Dirac (FD)
distribution at a common electronic temperature $T_{e}$ which is higher than
the lattice temperature $T$. Thus, after $\sim 100\ \text{fs}$, the
carriers and the lattice constitute two decoupled subsystems, most of the
energy remaining stored within the former ($T_{e}>T$).

If $n$ represents the conduction band carrier density, the balance equation
for $e$-$h$ pairs can be written as\cite{vandriel87}
\begin{equation}
   \frac{\partial n}{\partial t} + \nabla\cdot{\textbf J} = G + R\ ,
   \label{a}
\end{equation}
where $\textbf{J}$ is the current density, $G$ the carrier generation rate
resulting from the absorption of photons, and $R$ the net recombination rate
given by
\begin{equation}
   R = -\gamma n^{3} + \delta(T_{e}) n\ ,
\label{b}
\end{equation}
with $\gamma$ the Auger recombination coefficient and $\delta(T_{e})$ the impact
ionization coefficient. The carrier generation rate $G$ is the sum of one-
and two-photon processes, $G=G_{1}+G_{2}$, and can be written as
\begin{equation}
   G = \frac{(1-\Gamma)\alpha I(z,t)}{h\nu} + \frac{(1-\Gamma)^{2}\beta I^{2}(z,t)}{2h\nu}\ ,
\label{c}
\end{equation}
$(1-\Gamma)I(z,t)$ being the intensity of the laser at depth $z$ below the
surface, $\Gamma$ the reflection coefficient at the surface, and $\alpha$ and
$\beta$ the one-photon and two-photon interband absorption coefficients,
respectively. The absorption proceeds according to the Beer-Lambert law
\begin{equation}
   \frac{\partial I}{\partial z} = -(\alpha + \Theta n)I - \beta I^{2}\ ,
\label{d}
\end{equation}
where $\Theta$ represents the free-carrier absorption cross-section.

It is instructive to compare the two-photon to one-photon carrier generation
rates, that is the ratio
\begin{equation}
   \frac{G_{2}}{G_{1}} = (1-\Gamma)I(z,t)(\frac{\beta}{2\alpha})\ .
\label{e}
\end{equation}
If we choose $\lambda=308\ \text{nm}$, each photon carries an energy $h\nu=
4.03\ \text{eV}$, larger than the bandgap of silicon at 300~K,
$E_{G}\sim 1.12\ \text{eV}$.\cite{sze85} For Si at
300~K, $\Gamma =\Gamma_{\text{0}}\sim 0.59$,\cite{wood84}
$\alpha =\alpha_{\text{0}}\sim 1.5\times 10^{6}\
\text{cm}^{-1}$,\cite{wood84} and $\beta\sim 40\
\text{cm/GW}$.\cite{murayama94} For an irradiance as high as $10^{3}\
\text{GW/cm}^{2}$, a value at least an order of magnitude higher than the
ones considered in this work, Eq.~(\ref{e}) gives $G_{2}/G_{1}\lesssim\
0.01$. Moreover, it has been shown\cite{lietoila82} that free-carrier
absorption is negligible at short wavelengths, typically below 1000~nm.
Therefore, the one-photon interband transition is the dominant mechanism
for absorption in the present case and all other processes are ignored; as
noted above, this allows the photoemission of electrons to be neglected.

In practice, upon arrival at the surface, the $x$-$y$ position of a photon is
determined at random following a uniform probability distribution and
reflected, again at random, according to\cite{deunamuno89}
\begin{equation}
   \Gamma (T) = \Gamma_{\text{0}} + \Gamma_{R}\times T ,
\label{f}
\end{equation}
where $\Gamma_{R}=4\times 10^{-5}$. Upon melting of the
surface (see Sec.~\ref{subsec:below}), the reflectivity is assumed to be that
of \textit{l}-Si, i.e., $\Gamma_{l}=0.73$.\cite{wood84} If not reflected, the
photon is absorbed in agreement with Eq.~(\ref{d}), neglecting two-photon and
free-carrier absorption, at depth\cite{allen87}
\begin{equation}
   z = -\alpha ^{-1} \ln(\xi)\ ,
\label{g}
\end{equation}
where $\xi$ is a number generated from a uniform distribution; $\alpha$ is
the absorption coefficient, given by\cite{jellison82}
\begin{equation}
   \alpha (T) = \alpha_{\text{0}} \exp\left(\frac{T-\Theta_{\text{1}}}
   {\Theta_{\text{2}}}\right)\ ,
\label{h}
\end{equation}
where $\Theta_{\text{1}}=300\ \text{K}$ and $\Theta_{\text{2}}=4680\ \text{K}$.

The silicon atom closest to the absorption site, and which possesses at least
one valence electron, is then excited, i.e., its number of valence electrons
is decremented by one. An $e$-$h$ pair is created with an initial position
given by the $(x,y,z)$ coordinates of the excited atom. The relatively large
absorption coefficient at $\lambda=308\ \text{nm}$, combined with a depth of
60~nm for the simulation box, ensures that more than 99\% of the
energy is absorbed within the supercell.

The determination of the exact initial kinetic energy of each carrier
requires the knowledge of the allowed optical transitions. However, because
very fast carrier-carrier scattering leads to a FD distribution at
temperature $T_{e}$ within a few tens of fs, it is reasonable to assume the gas
to be in a state of an instantaneous quasi-equilibrium. Further, because
$T_{e}$ is relatively large, typically a few thousand degrees, the FD
distribution is well approximated by a MB distribution. Thus, as an electron
is promoted to the conduction band, its initial kinetic energy is calculated
according to a MB distribution at the current temperature $T_{e}$. The
problem of determining the set of optical transitions as well as the carrier
density and temperature-dependent chemical potential, required to
characterize a FD distribution, can thus be avoided. Using a Box-Muller
transformation,\cite{numrecipes} a MB distribution at $T_{e}$ is obtained by
calculating the electron initial kinetic energy, $E_{k_{i}}$, according to
\begin{equation}
   E_{k_{i}}=-k_{B} T_{e}\sum_{i=1}^{3}
   \ln(\zeta_{i})\cos^{2}(2\pi\xi_{i})\ ,
\end{equation}
where $\zeta_{i}$ and $\xi_{i}$ are random numbers taken from a uniform
distribution. The hole thus receives an initial kinetic energy equal to
($h\nu - E_{G} - E_{k_{i}}$). The relationship used for $E_{G}$ (in eV)
as a function of the lattice temperature $T$ and the carrier concentration
$n$ is\cite{vandriel87,pierret87}
\begin{equation}
   E_{G} = E_{G}^{\text{0}} - \frac{\kappa T^{2}}{(T + \Lambda)} -
\varsigma n^{\frac{1}{3}}\ ,
\end{equation}
where $E_{G}^{\text{0}}=1.17\ \text{eV}$, $\kappa=4.730\times\ 10^{-4}\ \text{eV/K}$,
$\Lambda=636\ \text{K}$ and $\varsigma=1.5\times 10^{-8}\ \text{eV\ cm}$.

\subsection{Relaxation processes\protect\\}
\label{subsec:relax}

A number of microscopic mechanisms allow the system to restore equilibrium
following irradiation by a laser pulse; these are carrier-phonon scattering,
carrier diffusion, impact ionization and Auger recombination.

Impact ionization is the process through which an $e$-$h$ pair is created
when an energetic electron collides with a valence electron: the former
transfers part of its kinetic energy to the latter which is promoted to the
conduction band. Because impact ionization most likely originates from the
generation of highly-energetic carriers through two-photon interband
transitions (negligible here; see Sec.~\ref{subsec:absorp}),\cite{schuler96}
it can be ignored.

In silicon, the dominant recombination mechanism is Auger\cite{landsberg91}
for which the characteristic recombination time, due to screening effects at
high carrier densities, is $\geq 6\ \text{ps}$.\cite{yoffa80} Because the carrier
density gradient, which develops at the surface following laser irradiation,
is taken into account, in our model, by assuming a unidirectional flow of electrons and
holes along the $z$ axis \textit{into} the bulk (see Sec.~\ref{subsec:diffusion}), the latter
have usually diffused out of the simulation box in a time less than 6~ps. As a result, Auger
recombination can also be neglected.

The important relaxation mechanisms, in the present context, are
carrier-phonon scattering and carrier diffusion, which we explicitly take
into account.

\subsubsection{Carrier-phonon scattering\protect\\}
\label{subsec:scatter}

If the carrier density $n$ remains below the critical
value\cite{stampfli94,silvestrelli96} $n_{c}\sim 10^{22}\ \text{cm}^{-3}$,
relaxation is thermal,\cite{linde97} i.e., the carriers relax by transferring
their kinetic energy to the lattice by the spontaneous emission of (mainly)
optical phonons in a characteristic time $\tau_{LO}\sim \text{1 ps}$
(also known as the carrier-phonon energy relaxation time $\tau_{E}$). The population of
optical phonons thus increases for approximately 1~ps, after which the latter relax by
emitting acoustic phonons in a characteristic time $\tau_{TA}\sim \text{10 ps}$. Eventually,
the energy is redistributed among all the vibrational modes of the lattice and a state of
quasi-equilibrium is reached, characterized by a Bose-Einstein distribution at temperature $T$.
The further evolution of the system can then be described through thermal
processes only as $T\longrightarrow T_{e}$, but $T$ is still higher than its
initial value.

There are several types of carrier-phonon scattering processes. It is
possible to identify the most important ones, as well as their associated
mean scattering rates. The dominant mechanisms for energetic electrons are
the $f$ and $g$ intervalley scattering transitions, which involve large
changes in momentum and thus acoustic or optical phonons with wave vectors
near the zone boundary.\cite{lundstrom90}

The probability $P_{if}$ of observing a given intervalley
transition, according to which a carrier in the valley $i$, with crystal
momentum \textbf{p}$_{i}$, scatters to a state with crystal momentum
\textbf{p}$_{f}$ in the valley $f$, is proportional to the square of the
intervalley deformation potential, $D_{if}$, which
characterizes the strength of the scattering.\cite{lundstrom90} The
probability can then be estimated as the weight associated with
$D_{if}^{2}$. The probabilities thus obtained are listed in
Table~\ref{tab:prob} and were calculated using the corresponding values for
$D_{if}$.\cite{fischetti88} As one can see, nearly 70\% of
phonons that scatter with electrons are optical. For holes, however, the main
mechanism is optical deformation potential scattering\cite{lundstrom90} and we
thus assume the holes to scatter only with optical phonons of 62.6~meV.\cite{lundstrom90}
Moreover, the energy of the carrier determines whether scattering involves
the absorption or the emission of a phonon. In Si, phonon emission is largely
dominant for carrier energies exceeding $E_{s}\sim \text{50 meV}$, while
phonon absorption dominates below $E_{s}$.\cite{lundstrom90}

To determine the rate at which the energy is transferred from the electronic
to the ionic degrees of freedom, the mean carrier-phonon scattering rate as a
function of carrier energy, $1/\tau^{\text{0}}_{cp}$, is also needed.
This quantity was calculated in Si at room temperature using the full
band-structure of the material by Fischetti~\textit{et al.}\cite{fischetti88}
A fit to their data was embedded in our computer model; variations with
temperature were ignored.

Because of screening effects, however, the observed carrier-phonon scattering rate
decreases for carrier densities $n\gtrsim 10^{21}\ \text{cm}^{-3}$, as recently
observed by Sjodin \textit{et al.};\cite{sjodin98} their results confirmed theoretical
predictions by Yoffa.\cite{yoffa81} Screening effects were thus implemented in our model as
suggested by Yoffa, with the new carrier-phonon scattering rate given by
\begin{equation}
\label{eq:screening}
   \tau_{cp}^{-1}(n) = \frac{1/\tau_{cp}^{\text{0}}}{(1 + (\frac{n}{n_{sc}})^{2})}\ ,
\end{equation}
where $n_{sc}\sim 1.2\times10^{21}\ \text{cm}^{-3}$ is the critical carrier
density to observe significant screening effects; $\tau_{cp}^{\text{0}}$ is the value of
$\tau_{cp}$ without screening. Finally, the probability of an electron suffering a collision with
a phonon was assumed to be Drude-like: the probability, $P_{s}$, that a carrier has scattered
with a phonon after a time $t$, where $t$ is the elapsed time since the last scattering event, is thus
\begin{equation}
   P_{s}(t)=1-\exp(-t/\tau_{cp})\ .
\label{eq:prob}
\end{equation}

In practice, we proceed as follows: after being generated, a carrier is given
an initial position and kinetic energy as described in
Sec.~\ref{subsec:absorp} and its motion proceeds until a scattering event
takes place. At each time step, the time elapsed since the last scattering
event is determined, $\tau_{cp}$ is computed and the
probability $P_{s}$ is determined from Eq.~(\ref{eq:prob}). If a scattering
event occurs, a phonon is either emitted or absorbed depending if the carrier
kinetic energy is greater or lower than $E_{s}$, respectively.

For an \textit{electron} scattering with a phonon, the intervalley process is
determined at random according to the probabilities of Table~\ref{tab:prob}.
A quantum of energy equal to $\hbar\omega_{\text{0}}$ is then given to
(phonon emission) or removed from (absorption) the lattice according to a
spatial Gaussian distribution in a radius of \text{5 \AA} from the carrier,
the latter value being an estimate of the carrier wave packet width based on
the uncertainty principle. Though a finite duration is associated with each
scattering event,\cite{bordone96} the latter are assumed to be instantaneous
for simplicity. Finally, if a phonon scattering event has occurred, the carrier
energy is updated.

\subsubsection{Carrier diffusion\protect\\}
\label{subsec:diffusion}

Carriers diffuse into the bulk as a result of the density gradient,
$\textbf{$\nabla$} n$, which is itself a consequence of the
exponentially-decreasing absorption with depth. The variations with $n$ of
the ambipolar diffusion coefficient $D$ and carrier mobilities, which
characterize the dynamics of carriers in semiconductors, have been the
subject of controversy; the available experimental data at densities
exceeding $10^{19}\ \text{cm}^{-3}$ are very scarce and concerns mostly
germanium.\cite{auston74,jamison76,moss81} In fact, for $n>10^{19}\
\text{cm}^{-3}$, it is not clear whether the ambipolar diffusion coefficient
decreases due to additional carrier-carrier scattering, as proposed by
Fletcher,\cite{fletcher57} or increases due to many-body quantum
effects.\cite{vandriel82} At the same time, various models have been
proposed\cite{vandriel82,kane92,velmre99} to explain the influence of
carrier-carrier scattering, as well as quantum effects, on $D$.

Recent experimental data\cite{linnros94,rosling94} for moderate
($10^{15}-10^{17}\ \text{cm}^{-3}$) carrier densities tend to support
Fletcher's model. In view of this, and the lack of experimental evidence on
many-body quantum effects for $n> 10^{19}\ \text{cm}^{-3}$, we adopt
Fletcher's model for $D$ implemented in our code using the expression
suggested by Berz \textit{et al.}\cite{berz79}

Thus, each $e$-$h$ pair is given an ambipolar diffusion coefficient which
depends upon the local carrier density and local temperature, and which is
updated at every time step, during which a carrier travels a distance $\Delta
z = \sqrt{D\Delta t}$ \textit{away} from the surface.

\section{Results and discussion\protect\\}
\label{sec:results}

All simulations were carried out assuming a laser pulse duration
$\tau_{L}=10\ \text{ps}$ and photon energy $h\nu=4.03\ \text{eV}$
($\lambda=308\ \text{nm}$). The pulse intensity has a temporal Gaussian
profile, but is spatially uniform. The fluence, $F$, was varied between 0.01
and 0.75~J/cm$^{2}$; corresponding intensities are in the range
1--75~GW/cm$^{2}$.

In Sec.~\ref{subsec:below}, we present our results for fluences of 0.01 to
0.20~J/cm$^{2}$. No matter removal is observed; a breathing mode of
the surface reveals the propagation of pressure waves in the solid. At and
above $F_{th}=0.25\ \text{J/cm}^{2}$, large pieces of molten material are
expelled from the surface; the mechanisms by which they are ejected are
examined in Sec.~\ref{subsec:above}.

\subsection{Below the threshold energy for ablation\protect\\}
\label{subsec:below}

\begin{figure}
\includegraphics*[width=7.5cm,height=5.8cm]{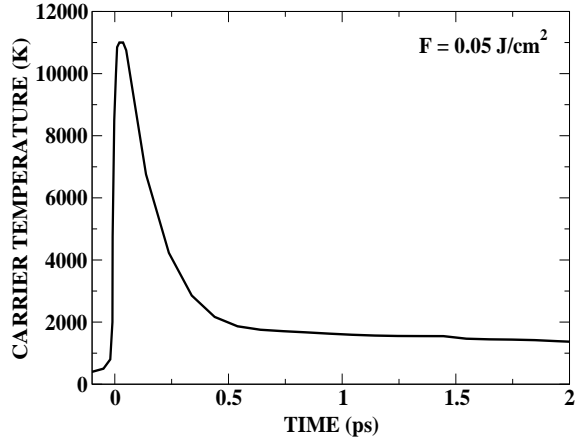}
\caption{Carrier temperature $T_{e}$ as a function of time for a fluence
$F=0.05\ \text{J/cm}^{2}$.}
\label{Te005}
\end{figure}

\begin{figure}
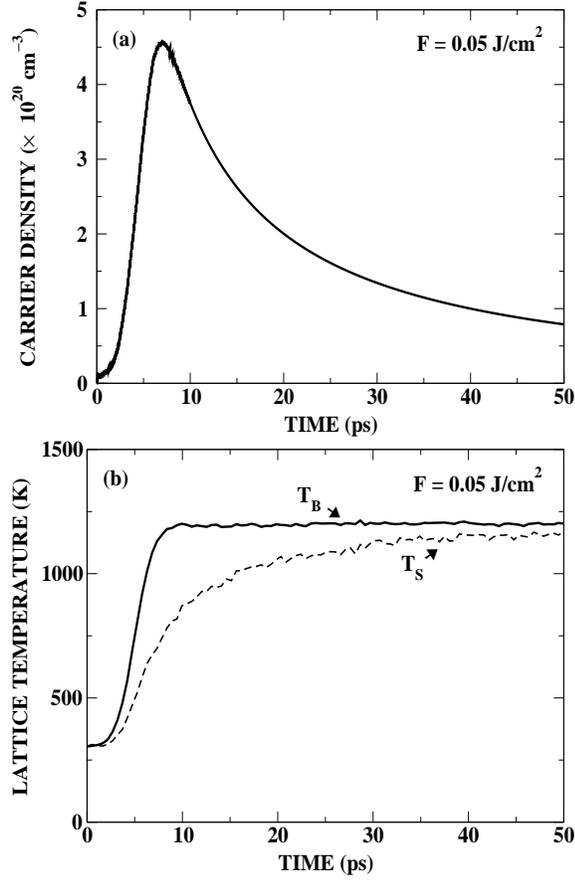

\includegraphics*[width=7.5cm,height=5.8cm]{N005}
\includegraphics*[width=7.5cm,height=5.8cm]{T005}
\caption{(a) Carrier density $n$ and (b) surface $T_{s}$ and bulk $T_B$
temperatures as a function of time for a fluence $F=0.05\ \text{J/cm}^{2}$.}
\label{TN005}
\end{figure}

Fig.~\ref{Te005} shows the carrier temperature $T_{e}$ during the first two
picoseconds for a fluence $F=0.05\ \text{J/cm}^{2}$; $T_{e}$ is obtained by
summing the kinetic energies of the electrons and holes in the conduction and
valence bands, respectively. The time evolution of the carrier density $n$
and the surface, $T_{s}$, and bulk, $T_{B}$, values of the
lattice temperature are shown in Fig.~\ref{TN005} for the same fluence; here,
$T_{s}$ is obtained by averaging over the first $\delta=1/\alpha\sim 7\ \text{nm}$
below the surface, where $\delta$ is the optical skin depth, while
$T_{B}$ represents an average over the rest of the supercell. The pulse starts at $t=0$
and the initial temperature of the target is $\sim 300\ \text{K}$. Upon arrival of the first photons,
$T_{e}$ rises rapidly to $\sim 11000\ \text{K}$ but $n$ remains low: the pulse provides sufficient
energy to create highly-energetic \textit{e-h} pairs while not significantly increasing
the number of carriers. The newly-generated hot electrons and holes promptly begin to
thermalize with the lattice: within $\sim 500\ \text{fs}$ to 1~ps, $T_{e}$ drops
to $\sim 1500\ \text{K}$; this indicates a very fast \textit{initial} cooling rate of
the carriers through phonon scattering. For delays $\gtrsim 1\ \text{ps}$, $T_{e}$
decreases at a slower rate, i.e., the rate at which energy is being transferred
from the electronic to the ionic degrees of freedom diminishes significantly. The
carrier-phonon energy relaxation time, $\tau_{E}$, is thus estimated to be a few hundred fs.
These observations are consistent with recent probing of ultrafast carrier dynamics in silicon
by Goldman \textit{et al.}\cite{goldman94} who reported a value of $\tau_{E}\sim 1\ \text{ps}$.
Equilibrium with the lattice is finally achieved after $\sim 10\ \text{ps}$ with
$T_{e}\longrightarrow \text{300 K}$.

The carrier density $n$ in Fig.~\ref{TN005}a reaches a peak at $t\sim 7\ \text{ps}$,
thus well after $T_{e}$ has passed its maximum; this is because the
\textit{e-h} plasma loses energy to the lattice faster than it gains energy
from the pulse. The decrease of $n$ for $t>7\ \text{ps}$ is due to diffusion
of the \textit{e-h} pairs away from the surface; moreover, the carrier
density remains well below the critical value, $n_{c}\sim 10^{22}\ \text{cm}^{-3}$,
ensuring that the TAM is appropriate.

The pulse generates a high number of electrons and holes which, in turn, heat
up the lattice; this is seen in Fig.~\ref{TN005}b. For $t>\tau_{L}$, $T_{B}$
reaches a plateau at $\sim 1250\ \text{K}$; this value is below the melting temperature
of c-Si, $T_{m}=1685\ \text{K}$.\cite{handbook} The surface temperature $T_{s}$, on
the other hand, continues to increase beyond $\tau_{L}$, approaching $T_{B}$ asymptotically.
Most importantly, $T_{s}<T_{B}$ for the whole duration of the simulation. This
clearly indicates that the maximum lattice temperature is located
\textit{beneath} the surface, a direct consequence of carriers diffusing into
the bulk as shown below.

\begin{figure}
\includegraphics*[width=7.5cm,height=5.8cm]{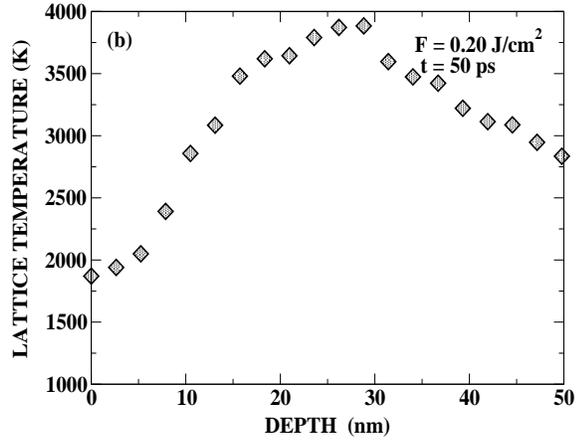}
\caption{Lattice temperature $T$ as a function of depth at $t=50\ \text{ps}$.
The fluence is $F=0.20\ \text{J/cm}^{2}$.}
\label{Tprof020}
\end{figure}

The fact that the maximum temperature is found below the surface can be
better appreciated from Fig.~\ref{Tprof020}, where the lattice temperature
$T$ is plotted as a function of depth $z$ at $t=50\ \text{ps}$ for a fluence
$F=0.20\ \text{J/cm}^{2}$; $z=0$ corresponds to the surface. The
temperature increases from a minimum of $\sim 2000\ \text{K}$ at the surface
to a maximum of $\sim 4000\ \text{K}$ at $z\sim 27\ \text{nm}$; the latter is
above the boiling point $T_{b}=2753\ \text{K}$, but below the critical
temperature, $T_{c}=5193\ \text{K}$, of silicon.\cite{handbook} For
$z\gtrsim 27\ \text{nm}$, $T$ decreases. Thus, the maximum temperature is not
occurring at the surface but, instead, at $z\sim 30\ \text{nm}$. These
observations can be explained as follows: the pulse intensity decreases
exponentially with depth and, consequently, more \textit{e-h} pairs are
created near the surface (see Sec.~\ref{subsec:absorp}). However, the time
interval between two consecutive carrier-phonon scattering events being
typically $\sim 10\ \text{fs}$ to a few ps, a carrier will diffuse, on
average, a few nm \textit{away} from the surface between two scattering
events. Thus, a large proportion of the energy is not released at the surface
but, rather, deeper in the bulk.

\begin{figure}
\includegraphics*[width=7.5cm,height=5.8cm]{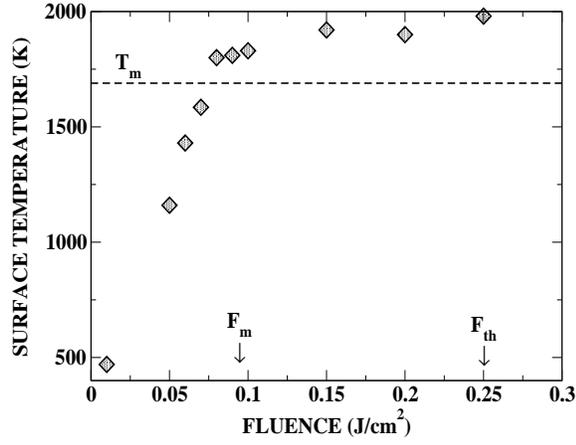}
\caption{Surface temperature $T_{s}$ as a function of laser fluence.}
\label{melt}
\end{figure}

Clearly, the temperature in Fig.~\ref{Tprof020} does not decay
exponentionally with depth, as one might have expected; rather, a steep
temperature gradient has developed at $z\lesssim 30\ \text{nm}$. This is a
clear indication of SSSH effects which are usually assumed to originate from
a vaporization process at the surface.\cite{singh90} However, it is important
to note that, in our simulations, no vaporization is observed; this
contrasts, for example, with the results of Zhigilei \textit{et al.} who
identified evaporation as the mechanism of matter removal below the threshold
energy for ablation.\cite{zhigilei97,zhigilei00} The absence of evaporation
in the present simulations is not surprising if one computes the flux (in
$\text{atoms}\ \text{m}^{-2}\ \text{s}^{-1}$) of atoms that desorb from the
surface, $\Phi_{e}$, given by\cite{ohring}
\begin{equation}
   \Phi_{e}=\frac{P_{e}(T_{s}^{e})}{\sqrt{2\pi Mk_{B}}T}\ ,
\label{eq:phi}
\end{equation}
where $M$ is the mass of a desorbed atom and $P_{e}$ is the pressure of the
vapor in equilibrium with the solid at a temperature $T_{s}^{e}$. For
$F\gtrsim 0.15\ \text{J/cm}^{2}$, $T_{s}\sim 2000\ \text{K}$ (see
Fig.~\ref{melt}), $P_{e}\sim\ 2.11\ \text{Pa}$,\cite{ohring} and we find
$\Phi_{e}\sim 2.3\times 10^{22}\ \text{atoms}\ \text{m}^{-2}\ \text{s}^{-1}$.
The surface of the supercell has an area $3\times 3\ \text{nm}^{2}=9\times
10^{-18}\ \text{m}^{2}$ and it is thus estimated that an atom desorbs every
$\sim 5\ \mu\text{s}$, which is much longer than the duration of our
simulations. Thus, the temperature profile in Fig.~\ref{Tprof020} cannot be
explained by an evaporation process; rather, it is the result of carriers
diffusing into the bulk.

\begin{figure}
\includegraphics*[width=7.5cm,height=5.8cm]{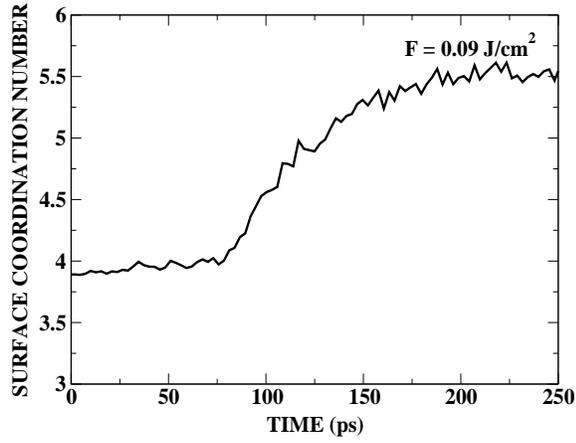}
\caption{Average surface coordination number $\rho_{\text{0}}$
as a function of time for a fluence $F=0.09\ \text{J/cm}^{2}$.}
\label{Cn009}
\end{figure}

Fig.~\ref{melt} shows the surface temperature $T_{s}$ as a function of
fluence. The threshold for melting, $F_{m}$, is thus estimated to be in the
range 0.05--0.1~GW/cm$^{2}$ since we must have $T_{s}\geq T_{m}$ to
observe melting. A more precise assessment of $F_{m}$ is obtained by
computing the coordination number averaged over the first few atomic layers
below the surface, $\rho_{\text{0}}$, as a function of time for various
fluences. The results for $F=0.09\ \text{J/cm}^{2}$ are depicted in
Fig.~\ref{Cn009}. A sudden change indicating the onset of melting is observed
at $t\sim 75\ \text{ps}$: from $\sim 3.88$ (a value slightly below that of
bulk c-Si, a consequence of including in the calculation undercoordinated
surface atoms), $\rho_{\text{0}}$ approaches the coordination number of $l$-Si,
i.e., $\sim 6$.\cite{wood84} Because no significant changes in the surface
coordination number are observed at lower fluences, the predicted threshold
is thus $F_{m}\sim 0.09\ \text{J/cm}^{2}$, melting taking place on a
characteristic time of a hundred ps, in good agreement with experimental
data.\cite{cavalleri99,bloembergen85,linde97}

\begin{figure}
\hspace{0.5cm} \includegraphics*[width=7.5cm,height=5.8cm]{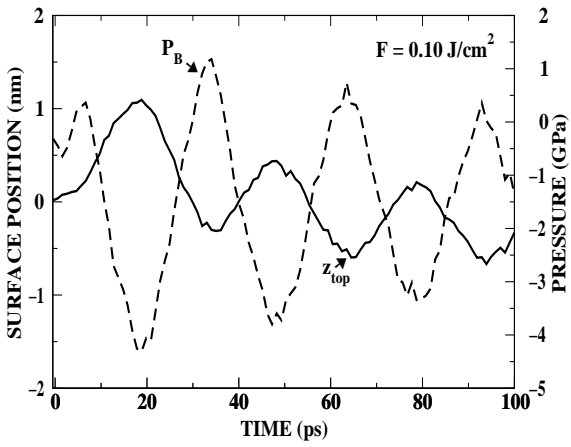}
\caption{Surface position $z_{top}$ and bulk pressure $P_{B}$ as a function
of time for a fluence $F=0.10\ \text{J/cm}^{2}$.}
\label{ztop010}
\end{figure}

The mechanisms by which the system restores internal equilibrium below
$F_{th}$ are revealed in Fig.~\ref{ztop010} where the surface position, i.e. the
$z$ coordinate of the topmost atomic layer, $z_{top}$, is
plotted with respect to time for $F=0.10\ \text{J/cm}^{2}$;
$z_{top}=0$ corresponds to the initial surface position. Also
plotted is the bulk pressure $P_{B}$, averaged over the
entire supercell. As one can see, oscillations with an amplitude of $\sim
1\ \text{nm}$ and a period $\sim 30\ \text{ps}$ are observed at the surface.
This breathing mode is due to the propagation of pressure waves in the solid;
the latter represent the elastic response of the material to the intense
heating by the pulse. Upon inspection of Fig.~\ref{ztop010}, $P_{B}$ and $z_{top}$
are clearly out of phase: because silicon expands when heated, $z_{top}$ initially
increases. At $t\sim 20\ \text{ps}$, $z_{top}$ is maximum and $P_{B}$ is minimum
($P_{B}<0$, tension); this is indicative of tensile stresses. In order to restore
internal equilibrium, the material subsequently enters a compressive phase and contracts, i.e.,
$z_{top}$ decreases. At $t\sim 35\ \text{ps}$, $z_{top}$ is minimum and $P_{B}$ is maximum
($P_{B}>0$, compression); the solid is then under compressive stresses. It is also apparent
from Fig.~\ref{ztop010} that melting occurs at the surface: the average value of $z_{top}$
decreases with time, an indication that the material is more compact. Indeed, molten silicon
is denser than crystalline silicon.\cite{handbook}

\begin{figure}
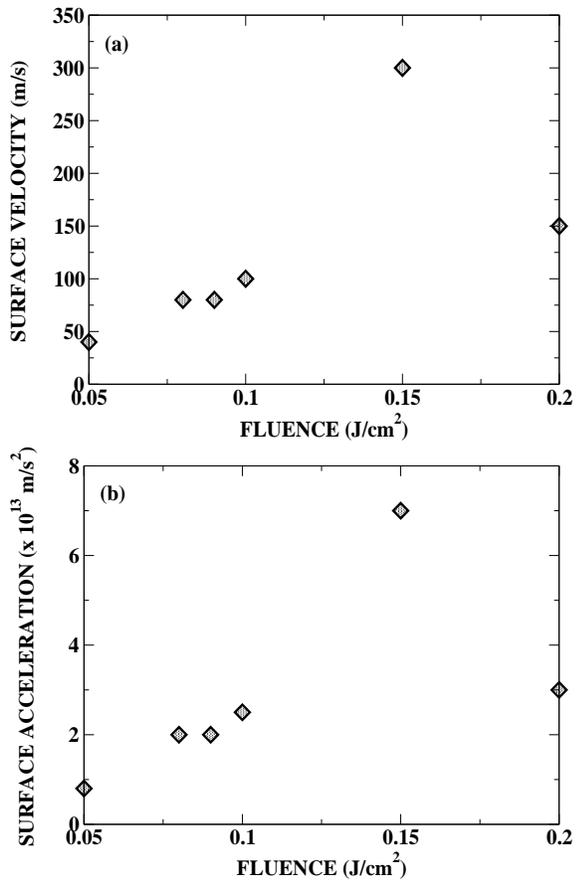

\includegraphics*[width=7.5cm,height=5.8cm]{vmax}
\includegraphics*[width=7.5cm,height=5.8cm]{amax}
\caption{Maximum velocity and maximum acceleration of the surface for
different fluences.}
\label{avmax}
\end{figure}

The surface velocity, $v_{top}$, and acceleration, $a_{top}$, computed from the data in
Fig.~\ref{ztop010}, are found to be $\sim 100\ \text{m/s}$ and $\sim 10^{13}\ \text{m/s}^{2}$,
respectively. These values are confirmed by noting that, upon heating, a surface undergoes
an expansion $\Delta z$ given by\cite{tam91}
\begin{equation}
   \Delta z=(1-\Gamma)F\alpha_{T}/\rho C\ ,
\label{eq:expand}
\end{equation}
where $\Gamma$ is the reflectivity, $\alpha_{T}$ is the linear
thermal expansion coefficient, $\rho$ is the density and $C$ is the specific
heat. Assuming $\Gamma\sim 0.59$, $\alpha_{T}\sim 2.6\times 10^{-6}\
\text{K}^{-1}$,\cite{silicon} $\rho\sim 2.33\ \text{g/cm}^{3}$,
$C\sim 0.713\ \text{J/gK}^{-1}$,\cite{silicon} and a fluence $F\sim 0.10\
\text{J/cm}^{-2}$ at 300~K, Eq.~(\ref{eq:expand}) yields an oscillation amplitude
$\Delta z\sim 0.6\ \text{nm}$, a value consistent with the results of Fig.~\ref{ztop010}.
With a pulse duration $\tau_{L}=10\ \text{ps}$, the velocity and acceleration of the
surface become $v_{top}\sim\Delta z/\tau_{L}\sim 60\ \text{m/s}$ and $a_{top}\sim\Delta
z/\tau_{L}^{2}\sim 10^{13}\ \text{m/s}^{2}$, thus in good agreement with the above
results. In addition, Fig.~\ref{avmax} gives the maximum value of
$v_{top}$ and $a_{top}$ as a function of laser fluence. A maximum is observed at
$F=0.15\ \text{J/cm}^{2}$. The subsequent decrease at $F=0.20\ \text{J/cm}^{2}$ might result
from a stronger damping of the oscillations due to a thicker molten layer at the surface.
Laser-induced oscillations at surfaces have found several applications. In dry laser
cleaning, for instance, the removal of contaminants, i.e., adsorbed
micron-sized particulates, is driven by momentum transfer: though of small
amplitude ($\sim 1\ \text{nm}$), the surface oscillations take place on a
very short time scale and thus exhibit very large accelerations.

\begin{figure}
\includegraphics*[width=7.5cm,height=5.8cm]{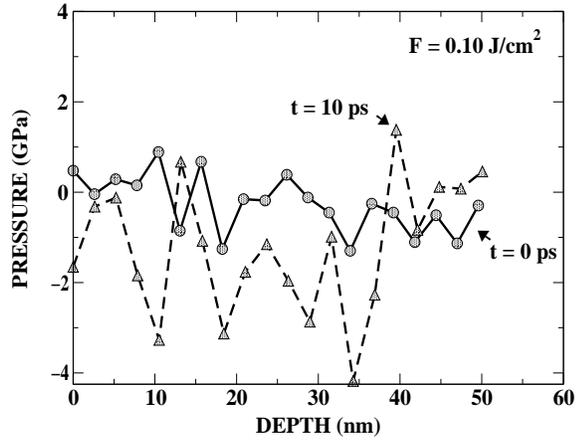}
\caption{Pressure $P$ as a function of depth at different times. The fluence
is $F=0.10\ \text{J/cm}^{2}$.}
\label{Pprof010}
\end{figure}

Finally, Fig.~\ref{Pprof010} shows the pressure $P$ as a function of depth
$z$ at $t=0$ and $t=10\ \text{ps}$; the fluence is $F=0.10\ \text{J/cm}^{2}$.
As expected, the pressure is zero throughout the material before laser
irradiation. By the end of the laser pulse, a pressure gradient has built up
at $35\lesssim z\lesssim 40$ nm; we show in Sec.~\ref{subsec:above} that the
latter is responsible for matter ejection at fluences $F\geq 0.25\ \text{J/cm}^{2}$.
Below the threshold energy for ablation, however, the pressure gradient is not high
enough to cause ejection of matter, and the system relaxes through the generation of pressure waves.

\subsection{Above the threshold energy for ablation\protect\\}
\label{subsec:above}

The threshold energy for ablation, $F_{th}$, is defined in this work as the
minimum fluence for the onset of matter removal. No evaporation is observed
(see Sec.~\ref{subsec:below}); rather, a collective ejection process is
obtained at all fluences above $F_{th}$. Here, $F_{th}=0.25\ \text{J/cm}^{2}$,
corresponding to a laser intensity $I_{th}=25\ \text{GW/cm}^{2}$.

\begin{figure}
\includegraphics*[width=7.5cm,height=5.8cm]{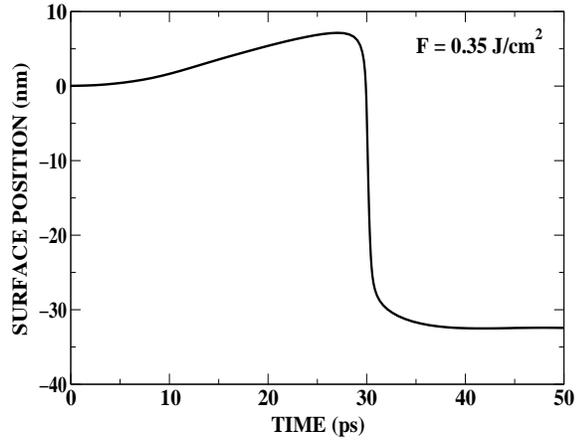}
\caption{Surface position $z_{top}$ as a function of time. The
fluence is $F=0.35\ \text{J/cm}^{2}$.}
\label{ztop035}
\end{figure}

Fig.~\ref{ztop035} shows the surface position, $z_{top}$, as a
function of time for a fluence $F=0.35\ \text{J/cm}^{2}$. As is the case
for $F<F_{th}$, the material initially expands upon heating. However this
does not result, this time, in a surface breathing mode. Instead, a dramatic
change occurs at $t\sim \text{30 ps}$: from a value of $\sim 7\ \text{nm}$,
$z_{top}$ drops to $\sim -35\ \text{nm}$ within a few ps. This indicates that
ablation has taken place over a depth $h\sim 35\ \text{nm}$, a
value consistent with experiment.\cite{cavalleri99} No significant change of
$h$ is observed at higher fluences.

\begin{figure}
\includegraphics*{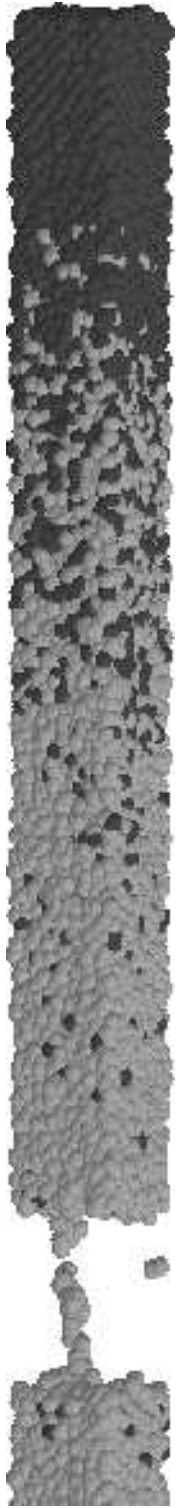}
\caption{Snapshot of the target at $t=33\ \text{ps}$ illustrating matter
removal. Atoms that have absorbed a photon are depicted in dark gray; others
are in light gray. The lateral and vertical dimensions are $\sim 3\ \text{nm}$
and $\sim 40\ \text{nm}$, respectively. The fluence is $F=0.35\ \text{J/cm}^{2}$.}
\label{snap1}
\end{figure}

A snapshot of the ablation process at $t=33\ \text{ps}$ for the same fluence
is given in Fig.~\ref{snap1}. The photons were incident from the top. The atoms
that have absorbed at least one photon are colored in dark gray, others are in
light gray. Evidently, most photons have been absorbed near the surface.

Two features should be noted: (i) Ablation is a \textit{bulk} phenomenon,
i.e., the ejection of matter originates in the bulk and {\em not} at the
surface as one might have expected. More specifically, ablation results in
the expulsion of a large piece --- a few tens of nm thick --- of molten
material from the surface with an initial axial velocity of $\sim 1000
\text{m/s}$. (ii) As opposed to a disordered network in the bulk, a few nm
thick crystalline layer is visible at the surface. The thickness of the
latter corresponds to the average distance, about 5~nm, carriers
created at the surface travel before suffering a first collision. For
$z\gtrsim 5\ \text{nm}$, the temperature rises rapidly and, as shown below,
the thickness of the layer removed during ablation depends upon the location
of the pressure build-up in the bulk.

\begin{figure}
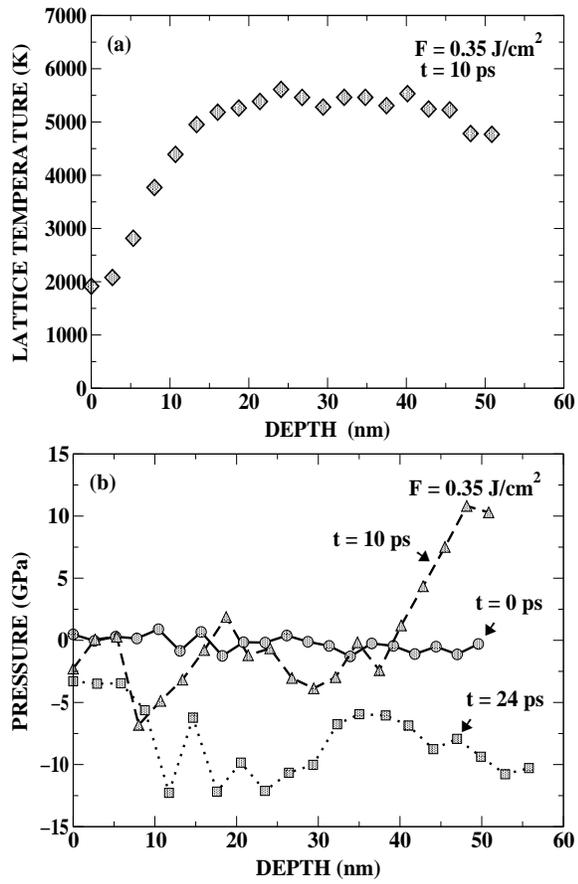

\includegraphics*[width=7.5cm,height=5.8cm]{Tprof035}
\includegraphics*[width=7.5cm,height=5.8cm]{Pprof035}
\caption{(a) Lattice temperature $T$ as a function of depth at $t=10\ \text{ps}$ and
(b) pressure $P$ at different times as a function of depth. The fluence is
$F=0.35\ \text{J/cm}^{2}$.}
\label{TPprof035}
\end{figure}

Insight into the mechanisms responsible for ablation is found in
Fig.~\ref{TPprof035} showing the pressure, $P$, and the lattice temperature,
$T$, as a function of depth for $F=0.35\ \text{J/cm}^{2}$.
A steep temperature gradient has developed at the surface by $t=10\
\text{ps}$, Fig.~\ref{TPprof035}a: from a value of $\sim 2000\ \text{K}$ at
$z=0$, $T$ rises rapidly to reach a plateau slightly above the critical
temperature, $T_{c}$, for $z\sim 30\ \text{nm}$. As shown in
Sec.~\ref{subsec:below}, the occurrence of a maximum temperature
\textit{below} the surface is not a consequence of evaporation from the
surface, nor does it result from a thermal diffusion process carried by
phonons. Rather, the energy is transported into the bulk by the electronic
degrees of freedom; this mechanism is, of course, absent in organic
solids.\cite{zhigilei00}

\begin{figure}
\begin{eqnarray}
\includegraphics*[width=2.25cm,height=7.5cm]{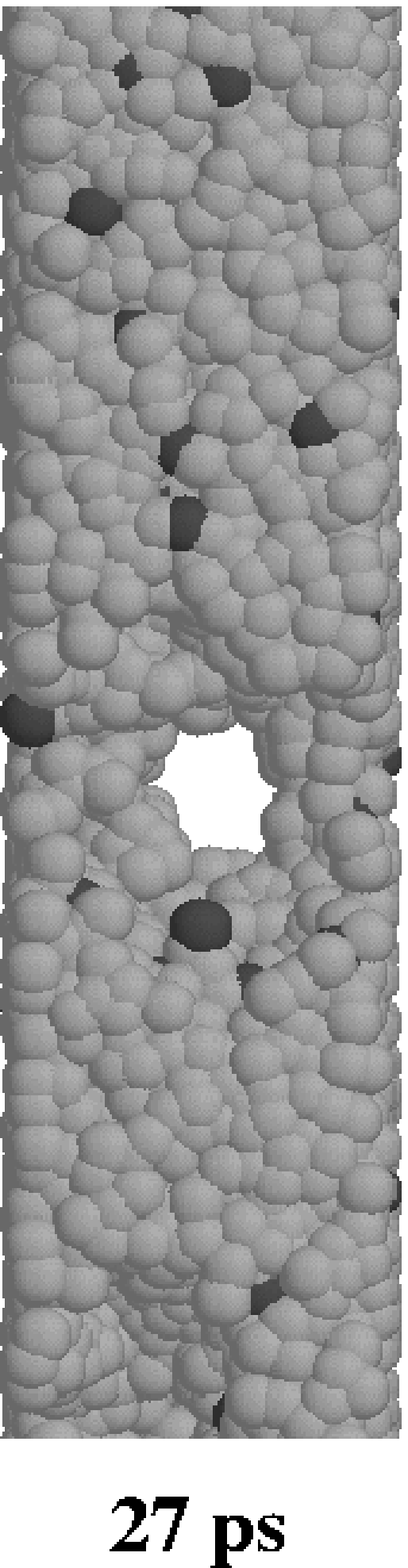} \nonumber
&
\hspace{0.375cm} \includegraphics*[width=2.25cm,height=7.5cm]{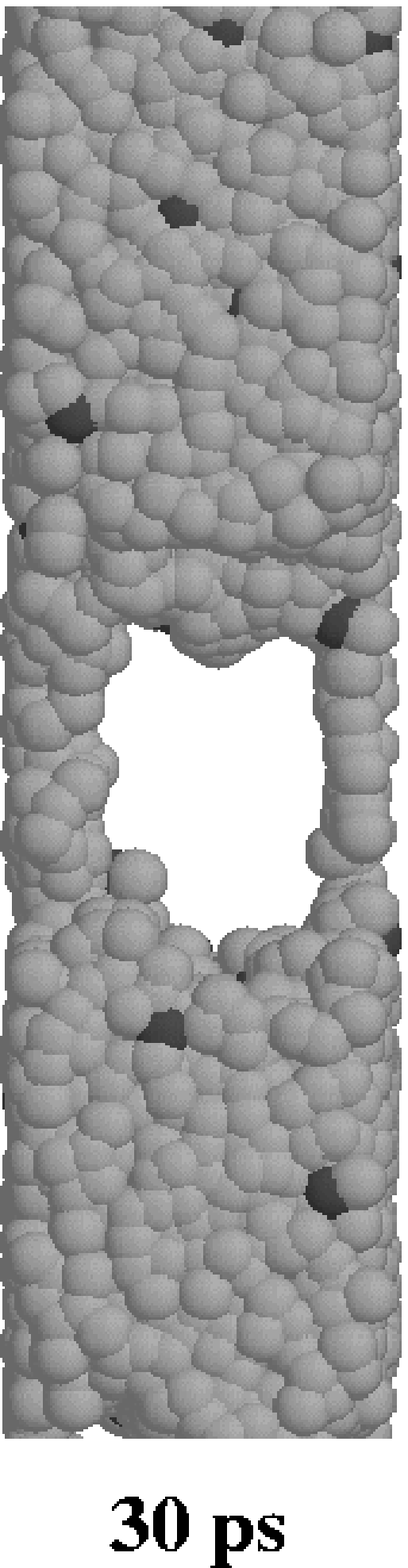} \nonumber
&
\hspace{0.375cm} \includegraphics*[width=2.25cm,height=7.475cm]{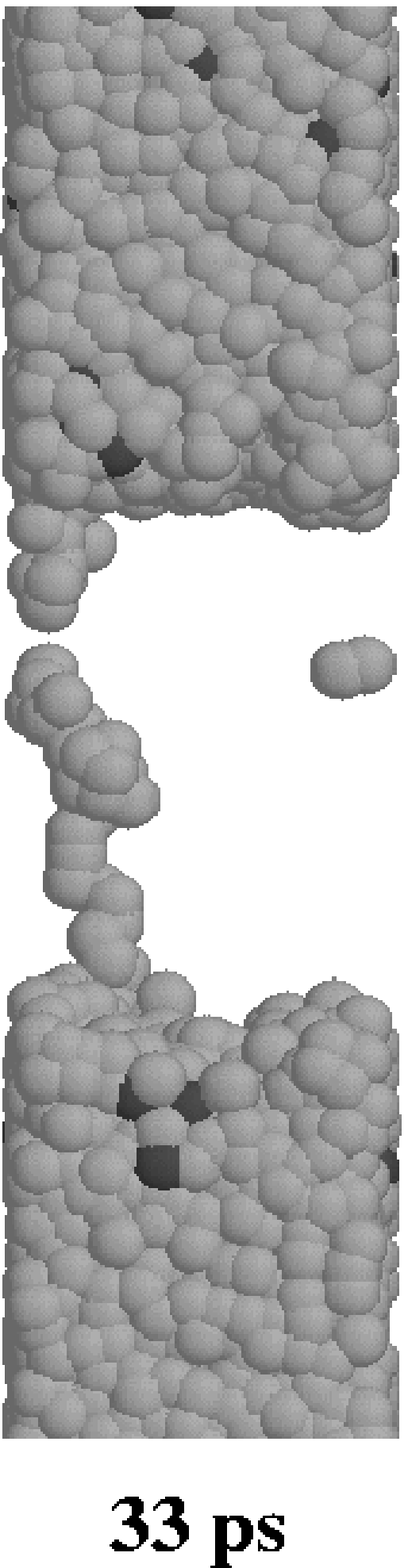} \nonumber
\end{eqnarray}
\caption{Snapshots of the subsurface region at a depth $30\lesssim z\lesssim
\text{40 nm}$. The surface (not shown here) is above. Atoms that have
absorbed a photon are depicted in dark gray; others are in light gray. The
strong tensile stresses are responsible for void nucleation beneath the
surface. The fluence is $F=0.35\ \text{J/cm}^{2}$.}

\label{snap2}
\end{figure}

Intense heating by the pulse leads to the thermal confinement of the
laser-deposited energy and, as shown in Fig.~\ref{TPprof035}a, the material
is heated far beyond its boiling point by the end of the pulse. At the same
time, the lattice cannot undergo sufficient thermal expansion and a strong
pressure ($> 0$) in excess of \text{10 GPa} develops at $35\lesssim z\lesssim
50\ \text{nm}$, Fig.~\ref{TPprof035}b. By $t\sim 24\ \text{ps}$ the latter
has relaxed, leaving important tensile stresses in the material; these are
responsible for the rapid expansion observed for $t\gtrsim 10\ \text{ps}$ in
Fig.~\ref{ztop035}. The pressure profile at $t=24\ \text{ps}$ in
Fig.~\ref{TPprof035}b also reveals a local minimum (in absolute value) at
$z\sim 35\ \text{nm}$. As a consequence, strong tensile forces of opposite
directions drive the material apart locally and nucleation of voids occur, as
depicted in Fig.~\ref{snap2}. The latter begins to form at $t=27\ \text{ps}$
and ablation follows a few ps later as the mechanical strength of the
material is exceeded.

\begin{figure}
\includegraphics*[width=7.5cm,height=5.8cm]{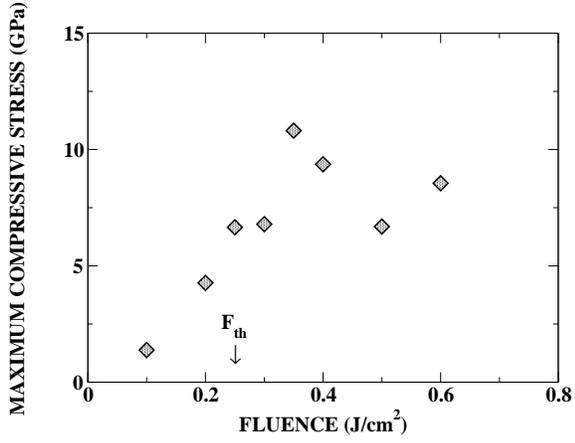}
\caption{Maximum compressive stress $P_{max}$ below the surface as a function
of fluence.}
\label{pmax}
\end{figure}

As seen above, the high compressive stress which develops within the
absorbing volume is responsible for the subsequent tensile forces; thus, it
is the driving force responsible for matter removal above $F_{th}$.
Fig.~\ref{pmax} shows the variation of the maximum compressive stress,
$P_{max}$, with laser fluence. Below the threshold for ablation, $P_{max}$
increases with $F$. Above $F_{th}$, however, $P_{max}$ saturates.

In a recent study of organic solids,\cite{zhigilei00} Zhigilei \textit{et
al.} reported the observation of void nucleation in the stress confinement
regime and attributed it to the build-up of a strong pressure gradient within
the absorbing volume. The condition of stress confinement is expressed by
$\tau_{L}<\tau_{S}\sim\delta/\nu_{S}$, where $\nu_{S}$ is the velocity of sound
in the material. For c-Si, $\nu_{S}\sim 9\times 10^{3}\ \text{m/s}$ and,\cite{handbook}
for a laser pulse at $\lambda=308\ \text{nm}$, $\tau_{S}\sim 1\ \text{ps}$. This
value is $<\tau_{L}$ by a factor of 10; conditions of stress confinement are thus
unlikely to develop in silicon following irradiation with 10~ps or longer pulses. The regime of
thermal confinement, on the other hand, is expressed by $\tau_{L}<\tau_{th}\sim\delta^{2}/D_{T}$,
where $D_{T}$ is the thermal diffusivity.\cite{zhigilei00} For silicon at 300~K,
$D_{T}\sim 0.86\ \text{cm}^{2}$/s;\cite{sze85} this corresponds to $\tau_{th}\sim 0.5\ \text{ps}$,
a value which is also about an order of magnitude lower than $\tau_{L}$. However, $D_{T}$
drops to $\sim 0.10\ \text{cm}^{2}$/s as $T$ reaches $\sim 1500\ \text{K}$,\cite{silicon} and it
follows that $\tau_{L}\sim\tau_{th}$: this suggests that conditions of thermal confinement
are likely to arise when silicon is heated beyond $T\sim 1500\ \text{K}$ in a typical time of 10~ps
or less. This phenomenon is responsible for the rapid heating of the material up to $T\sim T_{c}$
in Fig.~\ref{TPprof035}a. However, one should note that the presence of a
steep temperature gradient at the surface rules out the possibility of phase
explosion mechanisms which require that $\partial T/\partial z\sim\
0$.\cite{miotello95,kelly96} Consequently, ablation is here the result of
SSSH effects brought about by the thermal confinement of the energy.

Thus, the present simulations suggest that picosecond laser ablation of
silicon is driven by SSSH effects. However, a more precise assessment of the
matter removal mechanisms should involve the implementation of surface
effects; these have been observed in a study of hot carrier dynamics at a
Si(100) surface.\cite{goldman94} Accounting for surface effects is currently
under development.

\section{Concluding remarks}
\label{sec:conclusion}

A molecular-dynamics thermal annealing model (MADTAM) has been applied to the
study of laser ablation of silicon with picosecond pulses on a 100~ps
time scale. A detailed description of the thermal annealing model on a
microscopic scale has been embedded into a molecular-dynamics scheme; this
was accomplished by explicitly accounting for carrier-phonon scattering and
carrier diffusion.

The model predicts a melting threshold $F_{m}=0.09\ \text{J/cm}^{2}$ and an
ablation threshold $F_{th}=0.25\ \text{J/cm}^{2}$. Below $F_{th}$,
oscillations of the surface reveal the propagation of pressure waves in the
solid. Above $F_{th}$, important subsurface superheating (SSSH) effects are
responsible for matter removal: as a result of the thermal confinement of the
laser-deposited energy, the material is overheated to a temperature corresponding
to the critical point of silicon and a pressure gradient builds up a few tens of
nm \textit{beneath} the surface. Ejection of matter is initiated by the relaxation
of the strong compressive stresses after a few tens of ps, large pieces of molten
material being expelled from the surface with high axial velocities.

We are currently investigating the role of surface effects. These are
believed to be responsible for the formation of a surface space charge layer
which is expected to influence the carrier dynamics and to compete with
carrier diffusion.

\vspace{0.5cm}

\begin{acknowledgments}
PL wishes to thank L.~V. Zhigilei for the stimulating discussions. PL is also
grateful to Dietrich von~der~Linde, Henry~M. van~Driel as well as to
Massimo~V. Fischetti for their patience in answering questions. This work was
supported by the Natural Sciences and Engineering Research Council of Canada
and the ``Fonds pour la formation de chercheurs et l'aide \`{a} la
recherche'' of the Province of Qu\'{e}bec. We are indebted to the ``R\'eseau
qu\'eb\'ecois de calcul de haute performance'' (RQCHP) for generous allocations
of computer resources.
\end{acknowledgments}


\newpage

\vspace{1cm}

\begin{table}
\caption{Probabilities $P_{if}$ associated with the intervalley processes.}
\label{tab:prob}
\begin{center}
\begin{tabular}{llcc}
\text{} & $\hbar\omega_{\text{0}}$\ (meV) & $D_{if}$\ ($10^{8}\ \text{eV/cm}$) &
$P_{if}$ \\ \hline
$f_{\text{3}}$ & 62.6\ (LO,TO)\footnotemark[1] & 1.75\footnotemark[2] & 0.344 \\
$g_{\text{1}}$ & 44.3\ (TA)\footnotemark[2] & 1.18\footnotemark[2] & 0.156 \\
$g_{\text{2}}$ & 22.1\ (LA)\footnotemark[2] & 1.18\footnotemark[2] & 0.156 \\
$g_{\text{3}}$ & 62.6\ (LO)\footnotemark[1] & 1.75\footnotemark[2] & 0.344 \\
\text{Total} & \text{} & \text{} & 1.000 \\
\end{tabular}
\footnotetext[1]{Ref.~\onlinecite{sze85}.}
\footnotetext[2]{Ref.~\onlinecite{fischetti88}.}
\end{center}
\end{table}
%


\newpage

\vspace{1cm}



\begin{thebibliography}{99}

\bibitem{chase94}
L.~L. Chase, in \textit{Laser ablation: Principles and Applications}, edited by John~C. Miller, Springer Series in
Materials Science Vol.~25 (Springer-Verlag, Berlin, 1994).

\bibitem{haglund98}
R.~F. Haglund, Jr., \textit{Laser Ablation and Desorption}, edited by John~C. Miller and Richard~F. Haglund, Jr.,
Experimental Methods in the Physical Sciences Vol.~30 (Academic Press, San Diego, 1998).

\bibitem{chrisey}
D.~B. Chrisey and G.~K. Hubler, \textit{Pulsed Laser Deposition of Thin Films}, (John Wiley and Sons, New York, 1994).

\bibitem{belouet96}
C. Belouet, Appl. Surf. Sci. \textbf{96-98}, 630 (1996).

\bibitem{linde98}
K. Sokolowski-Tinten, J. Bialkowski, A. Cavalleri, D. von der Linde, A. Oparin, J. Meyer-ter-Vehn, and S.I. Anisimov,
Phys. Rev. Lett. \textbf{81}, 224 (1998).

\bibitem{linde00}
D. von der Linde and K. Sokolowski-Tinten, Appl. Surf. Sci. \textbf{154-155}, 1 (2000).

\bibitem{schmidt00}
V. Schmidt, W. Husinsky, and G. Betz, Phys. Rev. Lett. \textbf{85}, 3516 (2000).

\bibitem{jacques95}
A.~A. Oraevsky, S.~L. Jacques, and F.~K. Tittel, J. Appl. Phys. \textbf{78}, 1281 (1995)

\bibitem{zhigilei00}
L.~V. Zhigilei and B.~J. Garrison, J. Appl. Phys. \textbf{88}, 1281 (2000).

\bibitem{miotello95}
A. Miotello and R. Kelly, Appl. Phys. Lett. \textbf{67}, 3535 (1995).

\bibitem{kelly96}
R. Kelly and A. Miotello, Appl. Surf. Sci. \textbf{96-98}, 205 (1996).

\bibitem{dabby72}
F.~W. Dabby and U.~-~C. Paek, IEEE J. Quantum Opt. \textbf{QE-8}, 106 (1972).

\bibitem{bhattacharya91}
D. Bhattacharya, R.~K. Singh, and P.~H. Holloway, J. Appl. Phys. \textbf{70}, 5433 (1991).

\bibitem{singh90}
R.~K. Singh, D. Bhattacharya, and J. Narayan, Appl. Phys. Lett. \textbf{57}, 2022 (1990).

\bibitem{craciun98}
V. Craciun, D. Craciun, M.~C. Bunescu, C. Boulmer-Leborgne, and J. Hermann, Phys. Rev. B \textbf{58}, 6787 (1998).

\bibitem{craciun97}
V. Craciun, D. Craciun, M.~C. Bunescu, R. Dabu, and I.~W. Boyd, Appl. Surf. Sci. \textbf{109-110}, 354 (1997).

\bibitem{bennett95}
T.~D. Bennett, C.~P. Grigoropoulos, and D.~J. Krajnovich, J. Appl. Phys. \textbf{77}, 849 (1995).

\bibitem{wood84}
R.~F. Wood and G.~E. Jellison, in \textit{Pulsed Laser Processing of Semiconductors}, Semiconductors and
Semimetals Vol.~23 (Academic Press, Heidelberg, 1984), p. 166.

\bibitem{stillinger85}
F.~H. Stillinger and T.~A. Weber, Phys. Rev. B \textbf{31}, 5262 (1985).

\bibitem{lorazo00}
P. Lorazo, L.~J. Lewis, and M. Meunier, SPIE Proc. Series \textbf{3935}, 66 (2000).

\bibitem{lorazo00_2}
P. Lorazo, L.~J. Lewis, and M. Meunier, Appl. Surf. Sci. \textbf{168}, 276 (2000).

\bibitem{lietoila82}
A. Lietoila and J.~F. Gibbons, J. Appl. Phys. \textbf{53}, 3207 (1982).

\bibitem{vandriel87}
H.~M. van~Driel, Phys. Rev. B \textbf{35}, 8166 (1987).

\bibitem{pronko95}
P.~P. Pronko, S.~K. Dutta, D. Du, and R.~K. Singh, J. Appl. Phys. \textbf{78}, 6233 (1995).

\bibitem{seifert98}
N. Seifert and G. Betz, Appl. Surf. Sci. \textbf{133}, 189 (1998).

\bibitem{garrison85}
B.~J. Garrison and R. Srinivasan, J. Appl. Phys. \textbf{57}, 2909 (1985).

\bibitem{vertes95}
A. Bencsura and A. Vertes, Chem. Phys. Lett. \textbf{247}, 142 (1995).

\bibitem{campbell98}
R.~F.~W. Herrmann, J. Gerlach, and E.~E.~B. Campbell, Appl. Phys. A \textbf{66}, 35 (1998).

\bibitem{zhigilei97}
L.~V. Zhigilei, P.~B.~S. Kodali, and B.~J. Garrison, J. Phys. Chem. B \textbf{101}, 2028 (1997).

\bibitem{zhigilei98}
L.~V. Zhigilei, P.~B.~S. Kodali, and B.~J. Garrison, J. Phys. Chem. B \textbf{102}, 2845 (1998).

\bibitem{allen87}
M.~P. Allen and D.~J. Tildesley, \textit{Computer Simulation of Liquids}, (Oxford University Press, Oxford, 1987).

\bibitem{smith97}
R. Smith, M. Jakas, D. Ashworth, B. Oven, M. Bowyer, I. Charakov, and R. Webb, \textit{Atomic and Ion Collisions in Solids
and at Surfaces}, edited by R. Smith (Cambridge University Press, Cambridge, 1997).

\bibitem{zhigilei97_2}
L.~V. Zhigilei, P.~B.~S. Kodali, and B.~J. Garrison, Chem. Phys. Lett. \textbf{276}, 269 (1997).

\bibitem{cavalleri99}
A. Cavalleri, K. Sokolowski-Tinten, J. Bialkowski, M. Schreiner, and D. von~der~Linde, J. Appl. Phys. \textbf{85}, 3301
(1999).

\bibitem{cavalleri98}
A. Cavalleri, K. Sokolowski-Tinten, J. Bialkowski, and D. von~der~Linde, in \textit{Ultrafast Phenomena \textrm{XI}},
edited by T. Elsaesser, J.~G. Fujimoto, D.~A. Wiersma, and W. Zinth, Springer Series in Chemical Physics Vol. 63
(Springer, Heidelberg, 1998), p. 316.

\bibitem{linde97}
D. von~der~Linde, K. Sokolowski-Tinten, and J. Bialkowski, Appl. Surf. Sci. \textbf{109-110}, 1 (1997).

\bibitem{silvestrelli96}
P.~L. Silvestrelli, A. Alavi, M. Parrinello, and D. Frenkel, Phys. Rev. Lett. \textbf{77}, 3149 (1996).

\bibitem{linde00_2}
K. Sokolowski-Tinten and D. von~der~Linde, Phys. Rev. B \textbf{61}, 2643 (2000).

\bibitem{stampfli94}
P. Stampfli and K.~H. Bennemann, Phys. Rev. B \textbf{49}, 7299 (1994).

\bibitem{silvestrelli97}
P.~L. Silvestrelli, A. Alavi, M. Parrinello, and D. Frenkel, Phys. Rev. B \textbf{56}, 3806 (1997).

\bibitem{vanvechten79}
J.~A. Van~Vechten, R. Tsu, and F.~W. Saris, Phys. Lett. \textbf{74A}, 422 (1979).

\bibitem{murakami84}
K. Murakami and K. Masuda, in \textit{Semiconductors Probed by Ultrafast Laser Spectroscopy}, edited by R.~R. Alfano,
Vol. \textrm{II} (Academic Press, London, 1984).

\bibitem{schuler96}
D. von~der~Linde and H. {Sch\"{u}ler}, J. Opt. Soc. Am. B \textbf{13}, 216 (1996).

\bibitem{private}
D. von~der~Linde, private communication.

\bibitem{zhigilei99}
L.~V. Zhigilei and B.~J. Garrison, Mat. Res. Soc. Symp. Proc. \textbf{538}, 491 (1999).

\bibitem{handbook}
\textit{Handbook of Chemistry and Physics}, edited by D.~R. Lide (CRC Press, Boca Raton, 1993).

\bibitem{sze85}
S.~M. Sze, \textit{Physics of Semiconductor Devices}, (Wiley-Interscience, New York, 1969).

\bibitem{murayama94}
M. Murayama and T. Nakayama, Phys. Rev. B \textbf{49}, 5737 (1994).

\bibitem{deunamuno89}
S. De~Unamuno and E. Fogarassy, Appl. Surf. Sci. \textbf{36}, 1 (1989).

\bibitem{jellison82}
G.~E. {Jellison, Jr.,} and F.~A. Modine, Appl. Phys. Lett. \textbf{41}, 180 (1982).

\bibitem{numrecipes}
W.~H. Press, B.~P. Flannery, S.~A. Teukolsky, and W.~T. Vetterling, \textit{Numerical Recipes: the art of scientific computing}, (Cambridge
University Press, Cambridge, 1986).

\bibitem{pierret87}
R.~F. Pierret, \textit{Advanced Semiconductor Fundamentals}, edited by R.~F. Pierret and G.~W. Neudeck, Modular Series on
Solid State Devices Vol. \textrm{VI} (Addison-Wesley, Reading, 1987).

\bibitem{landsberg91}
P.~T. Landsberg, \textit{Recombination in Semiconductors}, (Cambridge University Press, Cambridge, 1991).

\bibitem{yoffa80}
E.~J. Yoffa, Phys. Rev. B \textbf{21}, 2415 (1980).

\bibitem{lundstrom90}
M. Lundstrom, \textit{Fundamentals of Carrier Transport}, edited by G.~W. Neudeck and R.~F. Pierret, Modular Series on
Solid State Devices Vol. \textrm{X} (Addison-Wesley, Reading, 1987).

\bibitem{fischetti88}
M.~V. Fischetti and S.~E. Laux, Phys. Rev. B \textbf{38}, 9721 (1988).

\bibitem{sjodin98}
T. Sjodin, H. Petek, and H.~L. Dai, Phys. Rev. Lett. \textbf{81}, 5664 (1998).

\bibitem{yoffa81}
E.~J. Yoffa, Phys. Rev. B \textbf{23}, 1909 (1981).

\bibitem{bordone96}
P. Bordone, D. Vasileska, and D.~K. Ferry, Phys. Rev. B \textbf{53}, 3846 (1996).

\bibitem{auston74}
D.~H. Auston and C.~V. Shank, Phys. Rev. Lett. \textbf{32}, 1120 (1974).

\bibitem{jamison76}
S.~A. Jamison, A.~V. Nurmikko, and H.~J. Gerritsen, Appl. Phys. Lett. \textbf{29}, 640 (1976).

\bibitem{moss81}
S.~C. Moss, J.~R. Lindle, M.~J. Mackey, and A.~L. Smirl, Appl. Phys. Lett. \textbf{39}, 227 (1981).

\bibitem{fletcher57}
N.~H. Fletcher, Proc. IRE \textbf{44}, 862 (1957).

\bibitem{vandriel82}
J.~F. Young and H.~M. van Driel, Phys. Rev. B \textbf{26}, 2147 (1982).

\bibitem{kane92}
D.~E. Kane and R.~M. Swanson, J. Appl. Phys. \textbf{72}, 5294 (1992).

\bibitem{velmre99}
E. Velmre and A. Udal, Physica Scripta \textbf{T79}, 193 (1999).

\bibitem{linnros94}
J. Linnros and V. Grivickas, Phys. Rev. B \textbf{50}, 16943 (1994).

\bibitem{rosling94}
M. Rosling, H. Bleichner, P. Jonsson, and E. Nordlander, J. Appl. Phys. \textbf{76}, 2855 (1994).

\bibitem{berz79}
F. Berz, R.~W. Cooper, and S. Fagg, Solid-State Electron. \textbf{22}, 293 (1979).

\bibitem{goldman94}
J.~R. Goldman and J.~A. Prybyla, Phys. Rev. Lett. \textbf{72}, 1364 (1994).

\bibitem{ohring}
M. Ohring, \textit{The Materials Science of Thin Films}, (Academic Press, Boston, 1992).

\bibitem{bloembergen85}
N. Bloembergen, Mat. Res. Soc. Symp. Proc. \textbf{51}, 3 (1985).

\bibitem{tam91}
A.~C. Tam, W.~P. Leung, W. Zapka, and W. Ziemlich, J. Appl. Phys. \textbf{71}, 3515 (1991).

\bibitem{silicon}
\textit{Properties of silicon}, (INSPEC, London, 1988).

\end{thebibliography}
\end{document}